# Study on radon and radium concentrations in drinking water in west region of Iran

Gh. Forozani[1] and Gh. Soori[2]

[1]Department of Physics, Bu_Ali Sina University, Hamedan-Iran
[2] Payam Noor University of Mashhad

**Abstract**

One of the most important characterizations of social health is existence the availability of safe drinking water. Since one of the sources of water contamination is nuclear contamination from radon gas, so in this research radon 222 concentration levels in water supplies in the Toyserkan (a region located in the west of Iran) is investigated. For measuring radon gas in water wells and springs Lucas chamber method is used. Review the results of these measurements that taken from 15th place show that, only five sites have radon concentrations above the limit dose. To reduce radon concentration, it is better to keep water in open pools in contact with air before the water is delivered to users.

## 1- Introduction

Two major chains of initial radioactive nuclear decay associated with the earth are Th-232 and U-238. Among the radioactive nuclear produced in the two mentioned chains, Radiom-226 is a product of the U-238 radioactive decay series with half-life about 1600 years that emits alpha particle is very important. Radon 222,the most stable isotope in its family which is used in radio therapy[1] and special properties as odourless, colourless, water-soluble and radioactive like-in other 20 isotopes, is the daughter of alpha disintegration from Ra-226 with a half-life about 92 hours, can then enter the human body through the respiratory. The total amount of annual effective dose of natural radioactive sources is 2.4 msv and1.2 msv nearly belongs to radon and its daughters[2]. Radon-222 and its decay products are the most important source of human exposures.

The second natural Radon isotope is Radon 219 which has a half-life 3.9 seconds and is produced from U 235 decay series. The third natural isotope located in the Th 232 decay chain is called Toron with half-life 55.6 seconds [3]. When the radon gas to be cooled below the freezing point it has a bright phosphoric color and in the lower temperature has yellow color. In normal temperature its color goes to be red orange.

The average concentration of radon in the atmosphere is about one molecule of radon in $10^{21}$ molecule of air, and it is estimated that every square mile of soil to a depth of 6 inches contains approximately one gram of radium which decays to radon 222 [4]. At least 50 percent of natural exposures are due to radon gas and its corresponding decay products. Many deaths from digestive and respiratory cancer are caused by radon exposure. [5,6]

According to recent information published by the United Nations Scientific Committee on effects of atomic radiation (UNSCERA 2008 report) inhalation of radon and its daughters, are the most important factors in happening natural human exposure from natural radioactive sources.

## 2- Radon in water and its hazards

E-mail: forozani@basu.ac.ir.

Once radon in water supplies reaches customers, it may produce human exposure via inhalation and direct digestion. Radon in water transfers into the air during showers, flushing toilets, washing dishes and washing clothes. The aerosols tend to deposit into the lungs where they release radiation that has been shown to increase the likelihood of lung cancer. Radon can also reach other body tissues through ingestion resulting in radiation exposure to the internal organs. Ingestion of radon is believed to increase the risk of stomach cancer.[7,8].

As stated one kinds of water contamination is due the existence of radon and this is very important in the quality of drinking water. It is remarkable that if the amount of radon in water is reduced only 40 to 50 percent before is delivered to users, it causes that the respiratory and gastrointestinal cancers reduced about 30 to 35 percent and in long term is very important for decreasing remedial cost. Considering the high rate of lung cancer in this region we had to investigate the amount of radon gas concentration in the water resources. The high rate of lung cancer in this region is our motivation to investigate the amount of radon gas concentration in the water resources.

### 3- Measurement of radon in water samples

The concentration of radon in underground water is more than of spring waters and also deep wells, however if the water is moving more and more the concentration of radon is expected to be further reduced. The investigated water samples are prepared with the outlet source from the lowest level of water depth with in 25 cm of surface water in the reverse pressure conditions. By keeping the samples in a cold place and at the least time the samples have been transported to laboratory. The samples were taken from 15 places that supply the drinking water and three samples of 330 ml have been taken from each site.

### 4- Radon measurement system

In this research to determine the concentration of radon in water the PRASSI (Portable Radon Gas Surveyor SILENA) system model 5S has been used. This system has suitable features to measure the concentration of radon gas in the water and weather. The most important characteristics of this system is the high sensitivity and short time response. The detection system has used the scintillation cell coated with Zn S(Ag) with 1830 $cm^3$ volume. This technique is one of the oldest and most reliable procedures for the measurement of radon. PRASSI pumping cycle operate at a constant rate with 3 litters per minute. In figure 1 the schematic of measuring PRASSI system is shown.

### 5- Measurement results

In this research, radon concentrations in water samples from 15 region of Tuyserkan is measured. The radon and radium concentration in each sample versus ($kBq/m^3$) and the annual effective dose of stomach, lung and whole body in terms of (mSv / year) have shown in table 1. In Figure 2 and 3 the histogram of radon and radium concentration of each sample are shown respectively. The results show that the amount of radon in five sites is more than $10(kBq/m^3)$. The annual effective dose for stomach and lung are calculated by using the results of Tayyeb *et.al*.[9].

Table 1. The radon and radium concentration of each sample $(kBq/m^3)$ and the annual effective dose of stomach, lung and whole body in terms of (mSv / year).

| Samples | Radon concentration $(\frac{kBq}{m^3})$ | Radium concentration $(\frac{kBq}{m^3})$ | Annual effective dose (mSv year$^{-1}$) | | |
|---|---|---|---|---|---|
| | | | Lung (inhalation) | Stomach(ingestion) | Whole body |
| Sample 1 | 5.915 | 0.34 | 15.545 | 16.503 | 32.048 |
| Sample 2 | 5.960 | - | 15.663 | 16.628 | 32.291 |
| Sample 3 | 5.971 | - | 15.692 | 16.660 | 32.352 |
| Sample 4 | 6.062 | 0.34 | 15.932 | 16.914 | 32.846 |
| Sample 5 | 6.582 | 0.90 | 17.299 | 18.365 | 35.664 |
| Sample 6 | 6.672 | 0.59 | 17.534 | 18.615 | 36.149 |
| Sample 7 | 6.674 | 0.49 | 17.542 | 18.624 | 36.166 |
| Sample 8 | 6.745 | - | 17.727 | 18.820 | 36.547 |
| Sample 9 | 6.778 | 0.4 | 17.812 | 18.910 | 36.722 |
| Sample 10 | 7.195 | 0.60 | 18.909 | 20.075 | 38.984 |
| Sample 11 | 9.142 | 1.42 | 24.026 | 25.503 | 49.529 |
| Sample 12 | 10.746 | 0.93 | 28.239 | 29.980 | 58.219 |
| Sample 13 | 10.942 | 0.96 | 30.529 | 28.757 | 59.286 |
| Sample 14 | 12.184 | 2.35 | 33.755 | 33.994 | 67.749 |
| Sample 15 | 14.360 | 3.14 | 34.738 | 40.064 | 74.802 |

## 6- Conclusions

Considering that about 50% of natural exposure of people is from radon gas and it is the leading cause of cancer patients suffering respiratory and gastrointestinal systems, and the highest percentage of radon that enter the human body is from drinking water and breathing, especially when is bathing, the measuring of radon gas in underground water in this region is done. The results of this research shows that radon gas concentration in five samples of water used by people have much more than the limit determined by the Environmental Protection Agency America $(10\frac{Bq}{Lit})$, but it is not very high and the amount of radon exist in water is not a serious hazard risk threaten the region. So for more attention to public health community and reducing the risks of radon gas, it is recommended that the drinking water must be kept in the open pools, or at least moved as a cascade to moving out radon gas.

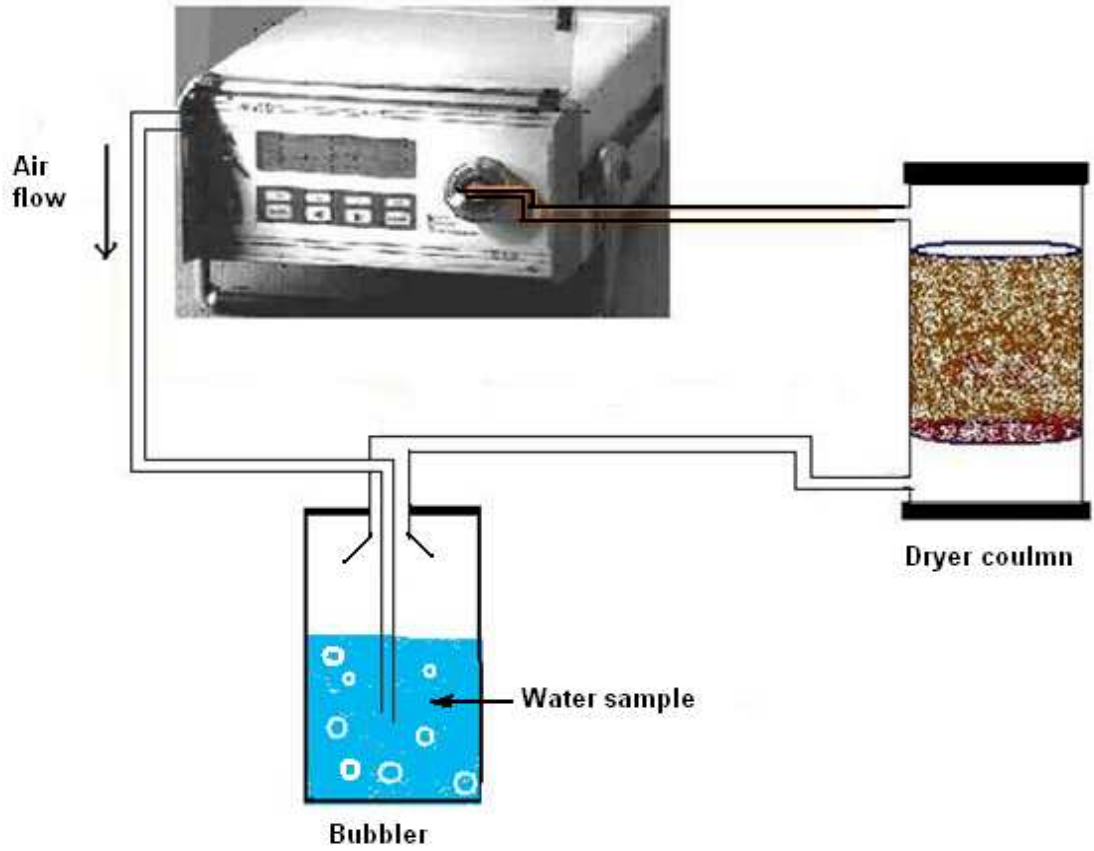

Figure 1. The schematic of measuring PRASSI system.

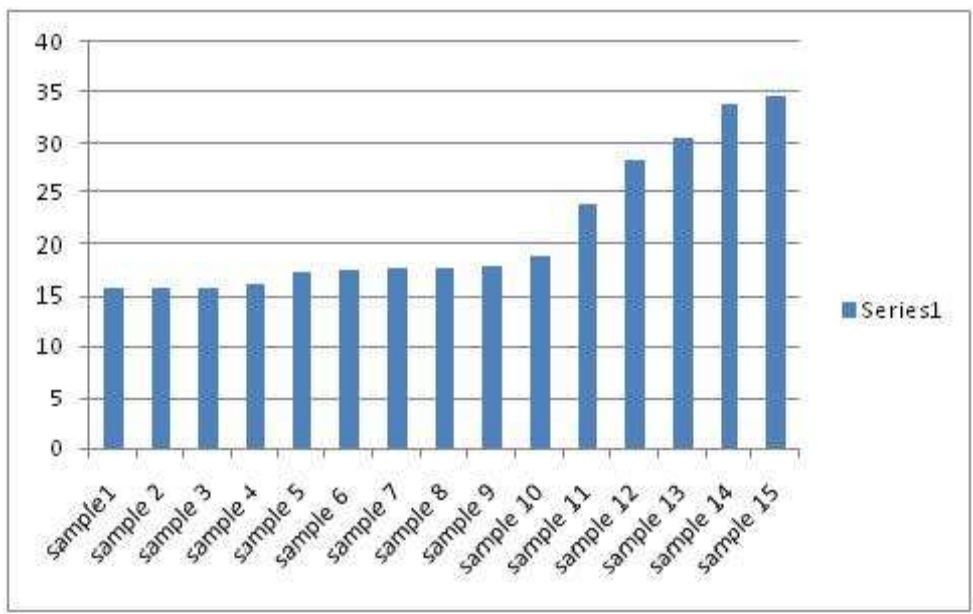

Figure 2. The histogram of radon concentration (Bq/Lit) of each sample.

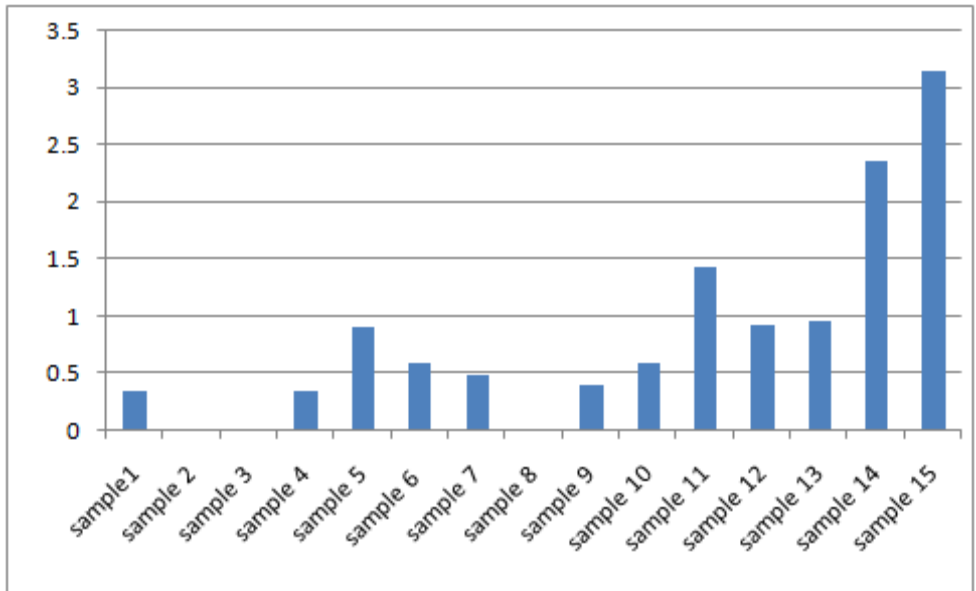

Figure 3. The histogram of radium concentration (Bq/Lit) of each sample.